\documentclass[11pt]{article}
\usepackage{amssymb}

\usepackage{amsmath}

\newcounter{resultnum}[section]

\newcounter{conclusionnum}[section]

\newcounter{conditionnum}[section]

\newcounter{conjecturenum}[section]

\newcounter{examplenum}[section]

\newcounter{exercisenum}[section]
\newtheorem{lemma}{Lemma}[section]

\newcounter{lemmanum}[section]

\newcounter{notationnum}[section]
\newtheorem{theorem}{Theorem}[section]

\newcounter{theoremnum}[section]

\newcounter{definitionnum}[section]

\newcounter{corollarynum}[section]

\newcounter{remarknum}[section]

\newcounter{propositionnum}[section]

\newcounter{acknowledgementnum}[section]

\newcounter{algorithmnum}[section]

\newcounter{axiomnum}[section]

\newcounter{casenum}[section]

\newcounter{claimnum}[section]

\newcounter{summarynum}[section]

\newcounter{problemnum}[section]

\begin{document}

\title{ Deformation Quantization of Nonholonomic \\
Almost K\"{a}hler Models and Einstein Gravity }
\date{December 22, 2007}
\author{ Sergiu I. Vacaru\thanks{%
sergiu$_{-}$vacaru@yahoo.com, svacaru@fields.utoronto.ca } \\
%EndAName
{\quad} \\
\textsl{The Fields Institute for Research in Mathematical Science} \\
\textsl{222 College Street, 2d Floor, } \textsl{Toronto \ M5T 3J1, Canada} }
\maketitle

\begin{abstract}
Nonholonomic  distributions and adapted frame
structures on (pseudo) Riemannian manifolds of even dimension are employed 
to build structures equivalent to almost K\"{a}hler geometry
and which allows to perform a Fedosov-like quantization of gravity.
The nonlinear connection formalism that was formally elaborated  for Lagrange and Finsler
geometry is implemented in classical and quantum Einstein gravity. 

\vskip10pt


\vskip5pt

\end{abstract}

\section{Introduction}

In recent years a set of important results in quantum gravity originated from Loop
Quantum Gravity (for reviews and discussion of results, including previous
canonical, topological, perturbative and other approaches, see Refs. \cite%
{rovelli,ashtekar}) that were developed in many cases as an alternative, see
discussion in \cite{smolin},  to stringy models of gravity (as general
references, see \cite{string1,string2,string3}).

Following different geometrical and nonlinear functional analytical 
methods,  a number of fundamental results were obtained based on 
deformation quantization
\cite{fedosov1,fedosov2,fedosov,konts1,konts2}.  Some
early attempts to apply the ideas and results from deformation
quantization to gravitational fields can be found,  for instance, in  
commutative and noncommutative self--dual gravity \cite{gc1,gc},
$W$--gravity \cite{castro1} and (recently) in  linearized gravity
\cite{qutaf}.  A geometric quantization formalism for Einstein's theory of gravity and its
gauge generalizations has not been elaborated yet.  One of the main differences between the
former canonical, perturbative  and loop geometry approaches
to quantum gravity and deformation quantization consists in
the fact that geometric quantization methods were elaborated
in curved spacetimes endowed with a certain symplectic, Poisson, or
almost K\"{a}hler structure, ...etc on the (co) tangent bundles, or, for
instance, for Lie algebroids. It was believed that symplectic
structures and generalizations cannot be introduced in a canonical
(unique) form in real Einstein manifolds.

In our recent work \ \cite{vrfg,vqgr1,vqgr2}, we proved that almost K\"{a}%
hler geometries can be obtained canonically for (pseudo)
Riemannian manifolds of even dimensions, including in Einstein
gravity, if the spacetime manifold is equipped  with a frame
structure and the associated nonlinear connection (N--connection).  
Briefly, a N--connection is defined in terms of  a nonholonomic
\footnote{In the 
mathematical and physics literature  different  
terminology  is used for this geometric property,  like nonintegrable,
non--holonomic and anholonomic --- all them are 
equivalent } 
distribution splitting the tangent space of the
spacetime manifold into conventional horizontal and vertical
subspaces. This means that a frame structure with mixed holonomic
and nonholonomic variables is being prescribed in the spacetime manifold.\footnote{%
One can use any system of reference and coordinates but certain
constructions are naturally adapted to some prescribed
nonholonomic structures} The N--connection components are
induced by certain off--diagonal elements of a metric
\footnote{in general, a four dimensional spacetime metric cannot
be diagonalized by coordinate transformations} which allows us to
construct a canonical almost K\"{a}hler geometry by deforming
nonholonomically the fundamental geometric objects of the  (pseudo)
Riemannian geometry.\footnote{%
it should be noted that physicists use the terms pseudo--Euclidean
and pseudo--Riemannian geometry but mathematicians are familiar,
for instance, with the term semi--Riemannian}

The goal of this paper is to provide a natural Fedosov
quantization procedure of  four dimensional (pseudo) Riemannian and Einstein
manifolds following the
N--connection formalism formally elaborated in Finsler and Lagrange geometry %
\cite{ma1987,ma} and generalized to nonholonomic manifolds and
quantum Lagrange--Finsler spaces in our partner work
\cite{vrfg,vqgr1,vqgr2,esv} (readers are recommended to see them
in advance).\footnote{ An integral and tensor calculus on
manifolds provided with an N--connection structure requires a more
sophisticate system of notations,  see details in the above-mentioned
partner works. We shall use the Einstein summation convention for local
expressions.} We shall develop our construction for metric
compatible connections with an effective torsion and a Neijenhuis
structure induced nonholonomically by the off--diagonal metric
components. 

The paper is organized as follows: In section 2 we show how
(pseudo) Riemannian spaces equipped with prescribed nonholonomic
distributions and nonlinear connection structures (i.e.
N--anholonomic manifolds) can be modeled equivalently in terms of almost
K\"{a}hler geometries. Section 3 is devoted to Fedosov's
quantization of such N--anholonomic spaces.  In section 4 we
briefly conclude that the nonholonomic frame method, when applied  to
the deformation quantization procedure,  provides a natural geometric
quantization method to any solution of the field equations in Einstein
gravity.

\section{Almost K\"{a}hler Models of Nonholnomic (pseu\-do) Ri\-emanni\-an
Spaces}

The aim of this section is to show how by prescribing corresponding
distributions (equivalently,  some classes of nonholonomic frame
structures) we can model a (pseudo) Riemannian manifold as an
almost K\"{a}hler space.

We consider a (pseudo) Riemann manifold $V^{2n},$ $\dim V^{2n}=2n,$ where $%
n\geq 2,$ of necessary smooth class.\footnote{%
For the Einstein gravity theory, $2n=4.$ In this paper, for
simplicity, we shall elaborate the geometric constructions
starting with smooth classical spacetime manifolds. Nevertheless,
 we have to keep in mind that certain physical situations may
request some more special geometric constructions when some
functions and/or their derivatives are not smooth in some
points/regions (for instance, black hole solutions). For such
cases, we have to "relax" the general smooth manifold condition
and postulate that we work with spacetimes and geometric objects
of "necessary" smooth class. } The local coordinates on $V^{2n}$
are labelled in the form $u^{\alpha }=(x^{i},y^{a}),$
or $u=(x,y), $ where indices run through the values $i,j,...=1,2,...n$ and $%
a,b,...=n+1,n+2,...,n+n,$ where $x^{i}$ and $y^{a}$ are called
 respectively the horizontal c (h) and
vertical (v) coordinates.

A nonlinear connection (N--connection) $\mathbf{N}$ on $V^{2n}$ is
defined as a Whitney sum (nonholonomic distribution)  on the
tangent bundle $TV^{2n},$
\begin{equation}
TV^{2n}=h(V^{2n})\oplus v(V^{2n}),  \label{whitney}
\end{equation}%
with a global splitting into conventional h-- and v--subspaces,
given locally by a set of
coefficients $N_{i}^{a}(x,y).$\footnote{%
defined with respect to a coordinate basis $\partial _{\alpha }=\partial
/\partial u^{\alpha }=(\partial _{i}=\partial /\partial x^{i},\partial
_{a}=\partial /\partial y^{a})$ and its dual $du^{\beta }=(dx^{j},dy^{b});$
here we note that the subclass of linear connections consists of a
particular case when $N_{i}^{a}=\Gamma _{\ bi}^{a}(x)y^{b}$} The curvature of a
N--connection is (by definition) just the Neijenhuis tensor
\begin{equation*}
\Omega _{ij}^{a}=\frac{\partial N_{i}^{a}}{\partial x^{j}}-\frac{\partial
N_{j}^{a}}{\partial x^{i}}+N_{i}^{b}\frac{\partial N_{j}^{a}}{\partial y^{b}}%
-N_{j}^{b}\frac{\partial N_{i}^{a}}{\partial y^{b}}.
\end{equation*}

In this work, the spacetimes will be modelled as N--anholonomic manifolds $%
\mathbf{V}^{2n},$ i.e. (pseudo) Riemanian manifolds with prescribed
nonholonomic distributions defining N--connection structures. For $n=2,$
fixing the Minkowski signature $(+---),$ we get a $2+2$ decomposition and
can develop a respective nonholonomic splitting formalism alternatively to
the so--called ADM (Arnowit, Deser and Misner) $\left( 3+1\right) $%
--decomposition (see, for instance, Ref. \cite{mtw}). We are going
to elaborate a canonical almost symplectic formalism which follows
naturally from certain type $\left( 2+2\right) $--decompositions.

Having prescribed on  $\mathbf{V}^{2n}$ a N--connection structure $\mathbf{%
N=\{}N_{j}^{a}\mathbf{\},}$ we can define a frame structure with
coefficients depending linearly on $N_{j}^{a},$ denoted $\mathbf{e}_{\nu }=(%
\mathbf{e}_{i},e_{a}),$ where
\begin{equation}
\mathbf{e}_{i}=\frac{\partial }{\partial x^{i}}-N_{i}^{a}(u)\frac{\partial }{%
\partial y^{a}}\mbox{ and
}e_{a}=\frac{\partial }{\partial y^{a}},  \label{dder}
\end{equation}%
and the dual frame (coframe) structure is $\mathbf{e}^{\mu }=(e^{i},\mathbf{e%
}^{a}),$ where
\begin{equation}
e^{i}=dx^{i}\mbox{ and }\mathbf{e}^{a}=dy^{a}+N_{i}^{a}(u)dx^{i},
\label{ddif}
\end{equation}%
satisfying nontrivial nonholonomy relations
\begin{equation}
\lbrack \mathbf{e}_{\alpha },\mathbf{e}_{\beta }]=\mathbf{e}_{\alpha }%
\mathbf{e}_{\beta }-\mathbf{e}_{\beta }\mathbf{e}_{\alpha }=W_{\alpha \beta
}^{\gamma }\mathbf{e}_{\gamma }  \label{anhrel}
\end{equation}%
with (antisymmetric) anholonomy coefficients $W_{ia}^{b}=\partial
_{a}N_{i}^{b}$ and $W_{ji}^{a}=\Omega _{ij}^{a}.$\footnote{%
We use boldface symbols for the spaces (geometric objects) provided with
N--connection structure (adapted to the h-- and v--splitting defined by (\ref%
{whitney}), we call them to be N--adapted). The N--adapted
tensors, vectors, forms etc are called respectively distinguished
tensors, vectors etc, in brief, d--tensors, d--vectors, d--forms
etc.}

Any metric $\mathbf{g=\{}g_{\alpha \beta }\mathbf{\}}$ on $\mathbf{V}^{2n},$
with symmetric coefficients $g_{\alpha \beta }$ defined with respect to a
coordinate dual basis $du^{\alpha }=(dx^{i},dy^{a}),$ can be represented
equivalently as d--tensor fields
\begin{eqnarray}
\mathbf{g} &=&g_{ij}(x,y)\ e^{i}\otimes e^{j}+h_{ab}(x,y)\ \mathbf{e}%
^{a}\otimes \ \mathbf{e}^{b},  \label{hvmetr} \\
&=&g_{ij}(x,y)e^{i}\otimes e^{j}+g_{a^{\prime }b^{\prime }}(x,y)\mathbf{%
\breve{e}}^{a^{\prime }}\otimes \ \mathbf{\breve{e}}^{b^{\prime }},
\label{hvmetr1}
\end{eqnarray}%
where \
\begin{equation}
h_{ab}(u)=g_{a^{\prime }b^{\prime }}(u)\mathbf{e}_{\ a}^{a^{\prime }}(u)%
\mathbf{e}_{\ b}^{b^{\prime }}(u)\mbox{ and }\mathbf{\breve{e}}^{a^{\prime
}}=\mathbf{e}_{\ a}^{a^{\prime }}(u)\mathbf{e}^{a}  \label{ftrn}
\end{equation}
are the v--components of the ''new'' N--adapted co--bases $\mathbf{\breve{e}}%
^{\mu }=(e^{i},\mathbf{\breve{e}}^{a^{\prime }}),$ being dual to  $\mathbf{%
\breve{e}}_{\nu }=(\mathbf{\breve{e}}_{i},e_{a^{\prime }}).$\footnote{%
 to get locally a diagonal Minkowski metric, we may consider (formally)
  that some local base vectors and corresponding coordinates
 are proportional to the complex imaginary unity "$i$" even in classical gravity
 we work with real (pseudo) Riemannian spaces}. With respect to ''v'' bases, the
respective coefficients $g_{ij}$ are equal to $g_{a^{\prime
}b^{\prime }}.$ In a particular case, we can consider that a metric $\mathbf{%
g=\{}g_{\alpha \beta }\mathbf{\}}$ is a solution of the Einstein equations
on $\mathbf{V}^{2n}.$

From the class of affine connections on $\mathbf{V}^{2n},$ one prefers to
work with N--adapted linear connections, called distinguished connections
(in brief, d--connections). A d--connection $\mathbf{D}=(hD;vD)=\{\mathbf{%
\Gamma }_{\beta \gamma }^{\alpha
}=(L_{jk}^{i},L_{bk}^{a};C_{jc}^{i},C_{bc}^{a})\},$ with local
coefficients computed with respect to (\ref{dder}) and
(\ref{ddif}), preserves under parallel transports the distribution
(\ref{whitney}).  For a metric compatible d--connection $\mathbf{D,}$  $\mathbf{D%
}_{\mathbf{X}}\mathbf{g}=0,$ for any d--vector $\mathbf{X.}$

We can define respectively the torsion and curvature tensors,
\begin{eqnarray}
\mathbf{T}(\mathbf{X},\mathbf{Y}) &\doteqdot &\mathbf{D}_{\mathbf{X}}\mathbf{%
Y}-\mathbf{D}_{\mathbf{Y}}\mathbf{X}-[\mathbf{X},\mathbf{Y}],  \label{ators}
\\
\mathbf{R}(\mathbf{X},\mathbf{Y})\mathbf{Z} &\doteqdot &\mathbf{D}_{\mathbf{X%
}}\mathbf{D}_{\mathbf{Y}}\mathbf{Z}-\mathbf{D}_{\mathbf{Y}}\mathbf{D}_{%
\mathbf{X}}\mathbf{Z}-\mathbf{D}_{[\mathbf{X},\mathbf{Y}]}\mathbf{Z},
\label{acurv}
\end{eqnarray}%
where the symbol ''$\doteqdot $'' states ''by definition'' and $[\mathbf{X},%
\mathbf{Y}]\doteqdot \mathbf{XY}-\mathbf{YX,}$ for any d--vectors $\mathbf{X}
$ and $\mathbf{Y}.$ With respect to fixed local bases $\mathbf{e}_{\alpha }$
and $\mathbf{e}^{\beta },$ the coefficients $\mathbf{T}=\{\mathbf{T}_{\
\beta \gamma }^{\alpha }\}$ and $\mathbf{R}=\{\mathbf{R}_{\ \beta \gamma
\tau }^{\alpha }\}$ can be computed by introducing $\mathbf{X}\rightarrow
\mathbf{e}_{\alpha },\mathbf{Y}\rightarrow \mathbf{e}_{\beta }\mathbf{,Z}%
\rightarrow \mathbf{e}_{\gamma }$ into respective formulas (\ref{ators}) and
(\ref{acurv}).

One uses three important geometric objects: the Ricci tensor, $Ric(\mathbf{D}%
)=\{\mathbf{R}_{\ \beta \gamma }\doteqdot \mathbf{R}_{\ \beta \gamma \alpha
}^{\alpha }\},$ the scalar curvature, $\ ^{s}R\doteqdot \mathbf{g}^{\alpha
\beta }\mathbf{R}_{\alpha \beta }$ ($\mathbf{g}^{\alpha \beta }$ being the
inverse matrix to $\mathbf{g}_{\alpha \beta }),$ and the Einstein tensor, $%
\mathbf{E}=\{\mathbf{E}_{\alpha \beta }\doteqdot \mathbf{R}_{\alpha \beta }-%
\frac{1}{2}\mathbf{g}_{\alpha \beta }\ ^{s}R\}.$  In Einstein
gravity, one works with the Levi Civita connection $\nabla =\{\
^{\shortmid }\Gamma _{\beta \gamma }^{\alpha }\}$
uniquely defined by a metric $\mathbf{g}$ in order to be metric compatible, $%
\nabla \mathbf{g=}0\mathbf{,}$ and torsionless, $\ ^{\shortmid }T_{\ \beta
\gamma }^{\alpha }=0.$ We emphasize that $\nabla $ is not a d--connection
because it does not preserve under parallel transports the splitting (\ref%
{whitney}) (that is why we do not use ''boldface'' letters).

For our purposes (to elaborate certain models of deformation quantization of
gravity), it is convenient to work with two classes of metric compatible
d--connections completely defined by a metric $\mathbf{g}$ and for a $2+2,$
in general, nonholonomic, splitting:

The first one is the so--called normal (in some cases, it is called also canonical)
 d--connection $\widehat{\mathbf{D}}%
=(h\widehat{D};v\widehat{D})=$ $\{\widehat{\mathbf{\Gamma }}_{\beta \gamma
}^{\alpha }\},$ or $h\widehat{D}=\{\widehat{L}_{jk}^{i}\}$ and $v\widehat{D}%
=\{\widehat{C}_{jc}^{i}\},$\footnote{%
It should be noted that on spaces of arbitrary dimension $n+m,m\neq n,$ we
have $\mathbf{D}=(hD;vD)=\{\mathbf{\Gamma }_{\beta \gamma }^{\alpha
}=(L_{jk}^{i},L_{bk}^{a};C_{jc}^{i},C_{bc}^{a})\};$ in such cases we can not
identify $L_{jk}^{i}$ with $L_{bk}^{a}$ and $C_{jc}^{i}$ with $C_{bc}^{a}.$
On a tangent bundle $TM,$ the h- and v--indices \ can be identified \ in the
form ''$i\rightarrow a",$ where $i=1=a=n+1,$ $i=2=a=n+2,$ ... $i=n=a=n+n.$}
when with respect to the N--adapted bases (\ref{dder}) and (\ref{ddif}),
\begin{equation}
\widehat{L}_{\ jk}^{i}=\frac{1}{2}g^{ih}(\mathbf{e}_{k}g_{jh}+\mathbf{e}%
_{j}g_{kh}-\mathbf{e}_{h}g_{jk}),\ \widehat{C}_{\ bc}^{a}=\frac{1}{2}%
h^{ae}(e_{b}h_{ec}+e_{c}h_{eb}-e_{e}h_{bc}).  \label{cdcc}
\end{equation}%
This connection is uniquely defined by the metric structure\ (\ref{hvmetr})
to satisfy the conditions $\widehat{\mathbf{D}}_{\mathbf{X}}\mathbf{g}=0$
and $\widehat{T}_{jk}^{i}=0$ and $\widehat{T}_{bc}^{a}=0$ (when torsion
vanishes in the h-- and v--subspaces); we note that, in general, the torsion
$\widehat{\mathbf{T}}_{\ \beta \gamma }^{\alpha }$ has nontrivial torsion
components
\begin{equation}
\widehat{T}_{jc}^{i}=\widehat{C}_{\ jc}^{i},\widehat{T}_{ij}^{a}=\Omega
_{ij}^{a},\widehat{T}_{ib}^{a}=e_{b}N_{i}^{a}-\widehat{L}_{\ bi}^{a},
\label{cdtors}
\end{equation}%
which are induced by respective nonholonomic deformations and also
completely defined by the N--connection and d--metric coefficients.

The second preferred metric compatible d--connection $\ ^{K}\mathbf{D}$ is
the same $\widehat{\mathbf{D}}$ but with the coefficients re--adapted with
respect to the bases $\mathbf{\breve{e}}_{\nu ^{\prime }}=(\delta
_{i^{\prime }}^{i}\mathbf{\breve{e}}_{i},e_{a^{\prime }})$ and $\mathbf{%
\breve{e}}^{\mu ^{\prime }}=(e^{i^{\prime }}=\delta _{i}^{i^{\prime }}e^{i},%
\mathbf{\breve{e}}^{a^{\prime }})$ (we use different symbols because we
consider a different N--adapted frame with redefined local coefficients of
N--connection, $\check{N}_{i}^{a^{\prime }}(u)=\mathbf{e}_{\ a}^{a^{\prime
}}(u)N_{i}^{a}(u),$ which defines a different nonholonomic distribution for
fixed structures $\mathbf{e}_{\ a}^{a^{\prime }}(u)$ and $N_{i}^{a}(u)$). In
general, we write $\mathbf{\breve{e}}_{\nu ^{\prime }}=\mathbf{\check{e}}%
_{v^{\prime }}^{\ v}(u)\mathbf{e}_{\nu }$ and $\mathbf{\breve{e}}^{\nu
^{\prime }}=\mathbf{\check{e}}_{\ v}^{v^{\prime }}(u)\mathbf{e}^{v},$ where
the matrix $\mathbf{\check{e}}_{\ v}^{v^{\prime }}$ is inverse to $\mathbf{%
\check{e}}_{v^{\prime }}^{\ v}$ defined correspondingly by $\delta
_{i^{\prime }}^{i},\delta _{a^{\prime }}^{a}$ and transforms with nontrivial
$\mathbf{e}_{\ a}^{a^{\prime }}$ stated by nonholonomic deformations (\ref%
{ftrn}). The N--adapted coefficients of $\ ^{K}\mathbf{D}$ can be expressed
through the N--adapted coefficients of $\widehat{\mathbf{D}}$ given by (\ref%
{cdcc}),
\begin{equation}
\ ^{K}\mathbf{\Gamma }_{\beta ^{\prime }\gamma ^{\prime }}^{\alpha ^{\prime
}}=\mathbf{\check{e}}_{\ \alpha }^{\alpha ^{\prime }}\mathbf{\check{e}}%
_{\beta ^{\prime }}^{\ \beta }\mathbf{\check{e}}_{\gamma ^{\prime }}^{\
\gamma }\mathbf{\Gamma }_{\beta \gamma }^{\alpha }+\mathbf{\check{e}}_{\
\alpha }^{\alpha ^{\prime }}\mathbf{e}_{\gamma }(\mathbf{\check{e}}_{\beta
^{\prime }}^{\ \alpha }).  \label{kdcon}
\end{equation}%
The torsion of this d--connection satisfies the condition
\begin{equation}
\ ^{K}\mathbf{T}_{\ \beta \gamma }^{\alpha }=(1/4)^{K}\mathbf{\Omega }_{\
\beta \gamma }^{\alpha },  \label{aux01}
\end{equation}%
where $^{K}\mathbf{\Omega }_{\ \beta \gamma }^{\alpha }$ are the N--adapted
coefficients of the Nijenhuis tensor
\begin{equation}
^{K}\mathbf{\Omega }(\mathbf{X},\mathbf{Y})\doteqdot \left[ \mathbf{\check{J}%
X,\check{J}Y}\right] -\mathbf{\check{J}}\left[ \mathbf{\check{J}X,Y}\right] -%
\mathbf{\check{J}}\left[ \mathbf{X,Y}\right] -\left[ \mathbf{X,Y}\right]
\label{neijt}
\end{equation}%
defined for the almost complex structure $\mathbf{\check{J}}:$ \
 $\mathbf{\check{J}}(\mathbf{\breve{e}}_{i})=-e_{i}$ and $\mathbf{%
\check{J}}(e_{i})=\mathbf{\breve{e}}_{i},$ where the superposition $\mathbf{%
\check{J}\circ \check{J}=-I,}$ for $\mathbf{I}$ being the unity matrix.%
\footnote{%
The formula (\ref{aux01}) was obtained in \cite{karabeg1,yano} for
arbitrary metric compatible affine connections on a manifold or
for lifts to tangent bundles. It holds true also for arbitrary
metric compatible d--connections. We note that in \cite{karabeg1}
there is a ''--'' sign because the authors define there the
Nijenhuis tensor with a different sign than in our case, see
(\ref{neijt}); they also use computations with respect to
coordinate bases.}

We note that representing the metric in the form (\ref{hvmetr1}) we are able
to define the almost symplectic structure in a canonical form,
\begin{equation}
\mathbf{\check{\theta}}=g_{ij}(x,y)\ \mathbf{\check{e}}^{i}\wedge e^{j},
\label{asymstr}
\end{equation}%
associated to $\mathbf{\check{J}}$ following formulas $\ \mathbf{\check{%
\theta}(X,Y)}\doteqdot \ \mathbf{g}\left( \mathbf{\check{J}X,Y}\right) ,$
which have the same h-- and v--components of metric. Defining an almost K%
\"{a}hler structure $\left( \mathbf{V}^{2n},\mathbf{\check{J},\check{\theta}}%
\right) $ for a N--anholonomic manifold $\mathbf{V}^{2n}$ enabled
with the N--connection $\mathbf{\check{N}}=\{
\check{N}_{j}^{a^{\prime }}\},$ we are able to treat $g_{ij}(x,y)$
as a generalized Lagrange metric but for a N--anholonomic
manifold, see discussions in \cite{vrfg,vqgr2}.

In Refs. \cite{ma1987,ma}, there are considered almost Hermitian
models for generalized Lagrange spaces on tangent bundles, when
the fiber N--adapted frame structure is holonomic. If we take an
arbitrary complex structure $J$
and define the almost symplectic form by the coefficients $\mathbf{\theta }%
_{\alpha \beta }=g_{\alpha \beta }(JX,Y),$ the constructions are
not adapted to the N--connection splitting. We have to work with
the Levi Civita connection $\nabla $ and perform not N--adapted
constructions.  In this way we do not have similar 
constructions as in Lagrange--Finsler
geometry (the last ones have the very important property that $%
g_{ij}=h_{ij}, $ for certain canonical N--connection structures)
and we cannot apply the N--connection formalism in order to
perform a canonical nonholonomic deformation quantification. In
this paper, considering a corresponding nonholonomic distribution
for the v--subspace, we are able to
elaborate constructions with closed symplectic forms resulting in almost K%
\"{a}hler geometry.

The nontrivial N--adapted coefficients of curvature $\widehat{\mathbf{R}}_{\
\beta \gamma \tau }^{\alpha }$ of $\widehat{\mathbf{D}}$ are
\begin{eqnarray}
\widehat{R}_{\ hjk}^{i} &=&\mathbf{e}_{k}\widehat{L}_{\ hj}^{i}-\mathbf{e}%
_{j}\widehat{L}_{\ hk}^{i}+\widehat{L}_{\ hj}^{m}\widehat{L}_{\ mk}^{i}-%
\widehat{L}_{\ hk}^{m}\widehat{L}_{\ mj}^{i}-\widehat{C}_{\ ha}^{i}\Omega
_{\ kj}^{a},  \label{cdcurv} \\
\widehat{P}_{\ jka}^{i} &=&e_{a}\widehat{L}_{\ jk}^{i}-\widehat{\mathbf{D}}%
_{k}\widehat{C}_{\ ja}^{i},\ \widehat{S}_{\ bcd}^{a}=e_{d}\widehat{C}_{\
bc}^{a}-e_{c}\widehat{C}_{\ bd}^{a}+\widehat{C}_{\ bc}^{e}\widehat{C}_{\
ed}^{a}-\widehat{C}_{\ bd}^{e}\widehat{C}_{\ ec}^{a}.  \notag
\end{eqnarray}%
For the d--connection $^{K}\mathbf{D,}$ with respect to $\mathbf{\breve{e}}%
_{\nu ^{\prime }}$ and $\mathbf{\breve{e}}^{\nu ^{\prime }},$ one has the
formulae%
\begin{equation}
\ ^{K}\mathbf{T}_{\beta ^{\prime }\gamma ^{\prime }}^{\alpha ^{\prime }}=%
\mathbf{\check{e}}_{\ \alpha }^{\alpha ^{\prime }}\mathbf{\check{e}}_{\beta
^{\prime }}^{\ \beta }\mathbf{\check{e}}_{\gamma ^{\prime }}^{\ \gamma }%
\widehat{\mathbf{T}}_{\beta \gamma }^{\alpha }\mbox{ and }\ ^{K}\mathbf{R}%
_{\ \beta ^{\prime }\gamma ^{\prime }\tau ^{\prime }}^{\alpha ^{\prime }}=%
\mathbf{\check{e}}_{\ \alpha }^{\alpha ^{\prime }}\mathbf{\check{e}}_{\beta
^{\prime }}^{\ \beta }\mathbf{\check{e}}_{\gamma ^{\prime }}^{\ \gamma }%
\mathbf{\check{e}}_{\tau ^{\prime }}^{\ \tau }\widehat{\mathbf{R}}_{\ \beta
\gamma \tau }^{\alpha },  \label{ktorscurv}
\end{equation}%
where $\widehat{\mathbf{T}}_{\beta \gamma }^{\alpha }$ and $\widehat{\mathbf{%
R}}_{\ \beta \gamma \tau }^{\alpha }$ have nontrivial coefficients given
respectively by formulas (\ref{cdtors}) and (\ref{cdcurv}).

It should be emphasized that we can work equivalently with both connections $%
\nabla $ and $\widehat{\mathbf{D}},$ because they are defined in a unique
form for the same metric structure (\ref{hvmetr}). All data computed for
one connection can be recomputed for another one by using the distorsion
tensor, $\widehat{\mathbf{Z}}_{\beta \gamma }^{\alpha },$ also uniquely
defined by the metric tensor $\mathbf{g}$ for the corresponding
N--connection splitting, when $\widehat{\mathbf{\Gamma }}_{\beta \gamma
}^{\alpha }=\ ^{\shortmid }\Gamma _{\beta \gamma }^{\alpha }+\mathbf{\ }%
\widehat{\mathbf{Z}}_{\beta \gamma }^{\alpha }.$

We conclude that a (semi) Riemannian N--anholonomic manifold provided with
a metric $\mathbf{g}$ and an N--connection $\mathbf{N}$ structure can be
equivalently described as a usual Riemannian manifold enabled with the
connection $\nabla (\mathbf{g})$ (in a form not adapted to the N--connection
structure) or as a Riemann--Cartan manifold with the torsion $\widehat{%
\mathbf{T}}(\mathbf{g})$ induced canonically by $\mathbf{g}$ and $\mathbf{N}$
(adapted to the N--connection structure). For an equivalent N--anholonomic
structure, we can work with $\ ^{K}\mathbf{T}(\mathbf{g}),$ induced
canonically by $\mathbf{g}$ and $\mathbf{\check{N}}$ and model the
constructions as for almost K\"{a}hler spaces.

\section{N--anholonomic Fedosov's Quantization}

Since we are interested in providing a natural deformation
quantization for (semi) Riemannian spaces equipped with
nonholonomic distributions, we have to reformulate the Fedosov
approach \cite{fedosov1,fedosov2,fedosov} for N--anholonomic
manifolds.

In this section, we perform the geometric constructions of Ref.
\cite{karabeg1} in a N--adapted
form, with respect to $\mathbf{\breve{e}}_{\nu }=(\mathbf{\breve{e}}%
_{i},e_{a^{\prime }})$ and $\mathbf{\breve{e}}^{\mu }=(e^{i},\mathbf{%
\breve{e}}^{a^{\prime }}),$  for a d--metric $\mathbf{g}$  (%
\ref{hvmetr1}) and d--connection $^{K}\mathbf{D}$ (\ref{kdcon}).
 Proofs of the results will be
omitted because they are completely similar to those for 
Lagrange--Finsler spaces \cite{vqgr2}, in N--adapted form, and to
those for metric affine connections, in coordinate (not
N--adapted) form, see \cite{karabeg1}.

Let us denote by $C^{\infty }(V)[[v]]$ the space of formal series in the 
variable $v$ with coefficients from a $C^{\infty }(V)$ on a Poisson manifold $%
(V,\{\cdot ,\cdot \}).$ An associative algebraic structure on
$C^{\infty }(V)[[v]],$ with a $v$--linear and $v$--adically
continuous star product
\begin{equation}
\ ^{1}f\ast \ ^{2}f=\sum\limits_{r=0}^{\infty }\ _{r}C(\ ^{1}f,\ ^{2}f)\
v^{r},  \label{starp}
\end{equation}%
is defined, where $\ _{r}C,r\geq 0,$ are bilinear operators on
$C^{\infty }(V),$  $\ _{0}C(\ ^{1}f,\ ^{2}f)=\ ^{1}f\ ^{2}f,$  $\
_{1}C(\ ^{1}f,\ ^{2}f)-\ _{1}C(\ ^{2}f,\ ^{1}f)=i\{\ ^{1}f,\
^{2}f\}$  and $i$ is the complex unity.

On $\mathbf{V}^{2n}$ enabled with a d--metric structure $\mathbf{g,}$ with
respect to $\mathbf{\breve{e}}_{\nu }=(\mathbf{\breve{e}}_{i},e_{a^{\prime
}}),$ we introduce the tensor $\ \mathbf{\check{\Lambda}}^{\alpha \beta
}\doteqdot \check{\theta}^{\alpha \beta }-i\ \mathbf{\check{g}}^{\alpha
\beta },$ where $\check{\theta}^{\alpha \beta }$ is the form (\ref{asymstr})
with ''up'' indices and $\ \mathbf{\check{g}}^{\alpha \beta }$ is the
inverse to $(g_{ij},g_{ab})$ stated by coefficients of (\ref{hvmetr1}). The
local coordinates on $\mathbf{V}^{2n}$ are parametrized  $%
u=\{u^{\alpha }\}$ and the local coordinates on $T_{u}\mathbf{V}^{2n}$ are
labelled $(u,z)=(u^{\alpha },z^{\beta }),$ where $z^{\beta }$ are the second
order fiber coordinates. We use the formal Wick product
\begin{equation}
a\circ b\ (z)\doteqdot \exp \left( i\frac{v}{2}\ \mathbf{\check{\Lambda}}%
^{\alpha \beta }\frac{\partial ^{2}}{\partial z^{\alpha }\partial
z_{[1]}^{\alpha }}\right) a(z)b(z_{[1]})\mid _{z=z_{[1]}},  \label{fpr}
\end{equation}%
for two elements $a$ and $b$ defined by formal series of type
\begin{equation}
a(v,z)=\sum\limits_{r\geq 0,|\overbrace{\alpha }|\geq 0}\ a_{r,\overbrace{%
\alpha }}(u)z^{\overbrace{\alpha }}\ v^{r},  \label{formser}
\end{equation}%
where $\overbrace{\alpha }$ is a multi--index, defining the formal Wick
algebra $\mathbf{\check{W}}_{u},$ for $u\in \mathbf{V}^{2n}$ associated
with the tangent space $T_{u}\mathbf{V}^{2n}.$

The fibre product (\ref{fpr}) is trivially extended to the space of $%
\mathbf{\check{W}}$--valued N--adapted differential forms $\ \mathcal{\check{%
W}}\otimes \Lambda $ by means of the usual exterior product of the scalar
forms $\Lambda ,$ where $\ \mathcal{\check{W}}$ denotes the sheaf of smooth
sections of $\mathbf{\check{W}.}$ For instance, in Ref. \cite{vqgr2}, we put
the left label $\mathcal{L}$ to similar values in order to emphasize that
the constructions are adapted to the canonical N--connection structure
induced by a regular effective Lagrangian; in this paper, we do not apply
such effective constructions and work with different types of d--metric,
N--connection and d--connection structures. All formulae presented below 
have certain analogies in Lagrange--Finsler geometry (this fact emphasizes the
generality of Fedosov's constructions) but in our case they will define
geometric constructions in nonholonomic (semi) Riemannian spaces, or in
Einstein spaces, and their almost K\"{a}hler models.

We can introduce a standard grading on $\Lambda \mathbf{,}$ denoted $%
\deg _{a},$and to introduce grading $\deg _{v},\deg _{s},\deg _{a}$ on $\
\mathcal{W}\otimes \Lambda $ defined on homogeneous elements $v,z^{\alpha },%
\mathbf{\check{e}}^{\alpha }$ as follows: $\deg _{v}(v)=1,$ $\deg
_{s}(z^{\alpha })=1,$ $\deg _{a}(\mathbf{\check{e}}^{\alpha })=1,$ and all
other gradings of the elements $v,z^{\alpha },\mathbf{\check{e}}^{\alpha }$
are set to zero. The product $\circ $ from (\ref{fpr}) on $\ \mathcal{\check{%
W}}\otimes \mathbf{\Lambda }$ is bi--graded, we write w.r.t the grading $%
Deg=2\deg _{v}+\deg _{s}$ and the grading $\deg _{a}.$

We extend $\ ^{K}\mathbf{D=\{}\ \ ^{K}\mathbf{\Gamma _{\alpha \beta
}^{\gamma }\}}$ (\ref{kdcon}) to an operator on $\mathcal{\check{W}}\otimes
\Lambda ,$
\begin{equation}
^{K}\mathbf{D}\left( a\otimes \lambda \right) \doteqdot \left( \mathbf{%
\check{e}}_{\alpha }(a)-u^{\beta }\ ^{K}\mathbf{\Gamma _{\alpha \beta
}^{\gamma }\ }^{z}\mathbf{\check{e}}_{\alpha }(a)\right) \otimes (\mathbf{%
\check{e}}^{\alpha }\wedge \lambda )+a\otimes d\lambda ,  \label{cdcop}
\end{equation}%
where $^{z}\mathbf{\check{e}}_{\alpha }$ is a similar to $\mathbf{\check{e}}%
_{\alpha }$ but depending on $z$--variables. The d--connection $\ \ ^{K}%
\mathbf{D}$ is a N--adapted $\deg _{a}$--graded derivation of the
distinguished algebra $\left( \mathcal{\check{W}}\otimes \mathbf{\Lambda
,\circ }\right) ,$ in brief, one call d--algebra (one follows from (\ref{fpr}%
) and (\ref{cdcop})).

The Fedosov operators $\check{\delta}$ and $\check{\delta}^{-1}$ are
defined, in our case, on$\ \ \mathcal{\check{W}}\otimes \mathbf{\Lambda }$ ,%
\begin{equation*}
\check{\delta}(a)=\ \mathbf{\check{e}}^{\alpha }\wedge \mathbf{\ }^{z}%
\mathbf{\check{e}}_{\alpha }(a)\mbox{ and }\ \ \check{\delta}%
^{-1}(a)=\left\{
\begin{array}{c}
\frac{i}{p+q}z^{\alpha }\ \mathbf{\check{e}}_{\alpha }(a),\mbox{ if }p+q>0,
\\
{\qquad 0},\mbox{ if }p=q=0,%
\end{array}%
\right.
\end{equation*}%
where $a\in \mathcal{\check{W}}\otimes \mathbf{\Lambda }$ is homogeneous
w.r.t. the grading $\deg _{s}$ and $\deg _{a}$ with $\deg _{s}(a)=p$ and $%
\deg _{a}(a)=q.$ Such operators define the formula $a=(\check{\delta}\
\check{\delta}^{-1}+ \check{\delta}^{-1}\ \check{\delta} +\sigma )(a),$
where $a\longmapsto \sigma (a)$ is the projection on the $(\deg _{s},\deg
_{a})$--bihomogeneous part of $a$ of degree zero, $\deg _{s}(a)=\deg
_{a}(a)=0.$ We note that $\check{\delta}$ is also a $\deg _{a}$--graded
derivation of the d--algebra $\left( \mathcal{\check{W}}\otimes \mathbf{%
\Lambda ,\circ }\right) .$

Using the extension of $^{K}\mathbf{D}$ to $\mathcal{\check{W}}\otimes
\mathbf{\Lambda ,}$ we construct the operators
\begin{eqnarray}
^{K}\mathcal{T}\ &\doteqdot &\frac{z^{\gamma }}{2}\ \check{\theta}_{\gamma
\tau }\ ^{K}\mathbf{T}_{\alpha \beta }^{\tau }(u)\ \mathbf{\check{e}}%
^{\alpha }\wedge \mathbf{\check{e}}^{\beta },  \label{at1} \\
\ ^{K}\mathcal{R} &\doteqdot &\frac{z^{\gamma }z^{\varphi }}{4}\ \check{%
\theta}_{\gamma \tau }\ ^{K}\mathbf{R}_{\ \varphi \alpha \beta }^{\tau }(u)\
\mathbf{\check{e}}^{\alpha }\wedge \mathbf{\check{e}}^{\beta },  \label{ac1}
\end{eqnarray}%
where the nontrivial coefficients of $\ ^{K}\mathbf{T}_{\alpha \beta }^{\tau
}$ and $\ ^{K}\mathbf{R}_{\ \varphi \alpha \beta }^{\tau }$ are computed as
in (\ref{ktorscurv}). One has the important formulae:
\begin{equation*}
\left[ ^{K}\mathbf{D},\check{\delta}\right] =\frac{i}{v}ad_{Wick}(^{K}%
\mathcal{T})\mbox{ and }\ ^{K}\mathbf{D}^{2}=-\frac{i}{v}ad_{Wick}(\ ^{K}%
\mathcal{R}),
\end{equation*}
where $[\cdot ,\cdot ]$ is the $\deg _{a}$--graded commutator of
endomorphisms of $\mathcal{\check{W}}\otimes \mathbf{\Lambda }$ and $%
ad_{Wick}$ is defined via the $\deg _{a}$--graded commutator in $\left(
\mathcal{\check{W}}\otimes \mathbf{\Lambda ,\circ }\right) .$

We reformulate three theorems \footnote{see Theorems 4.1, 4.2 and 4.3 in \cite{vqgr1},
for Lagrange--Finsler spaces; for effective Lagrange spaces, see \cite%
{vqgr2}; all such theorems are N--adapted generalizations of the
original Fedosov's results} and some fundamental properties of
Fedosov's d--operators for N--anholonomic (semi) Riemannian and
Einstein spaces:

\begin{theorem}
\label{th2}Any d--tensor (\ref{hvmetr1}) defines a flat canonical Fedosov
d--connec\-ti\-on
\begin{equation*}
\ ^{K}\mathcal{D}\doteqdot -\ \check{\delta}+\ ^{K}\mathbf{D}-\frac{i}{v}%
ad_{Wick}(r)
\end{equation*}%
satisfying the condition $\ \ ^{K}\mathcal{D}^{2}=0,$ where the unique
element $r\in $ $\mathcal{\check{W}}\otimes \mathbf{\Lambda ,}$ $\deg
_{a}(r)=1,$ $\check{\delta}^{-1}r=0,$ solves the equation
\begin{equation*}
\ \check{\delta}r=\ ^{K}\mathcal{T}\ +\ ^{K}\mathcal{R}+\ ^{K}\mathbf{D}r-%
\frac{i}{v}r\circ r
\end{equation*}%
and this element can be computed recursively with respect to the total
degree $Deg$ as follows:%
\begin{eqnarray*}
r^{(0)} &=&r^{(1)}=0,\ r^{(2)}=\check{\delta}^{-1}\ ^{K}\mathcal{T}\ , \\
r^{(3)} &=&\ \check{\delta}^{-1}\left( \ ^{K}\mathcal{R}+\ \ ^{K}\mathbf{D}%
r^{(2)}-\frac{i}{v}r^{(2)}\circ r^{(2)}\right) , \\
r^{(k+3)} &=&\ \ \check{\delta}^{-1}\left( \ ^{K}\mathbf{D}r^{(k+2)}-\frac{i%
}{v}\sum\limits_{l=0}^{k}r^{(l+2)}\circ r^{(l+2)}\right) ,k\geq 1.
\end{eqnarray*}%
where we denoted the $Deg$--homogeneous component of degree $k$ of an
element $a\in $ $\ \mathcal{\check{W}}\otimes \mathbf{\Lambda }$ by $%
a^{(k)}. $
\end{theorem}

\begin{theorem}
\label{th3}A star--product on the almost K\"{a}hler model of a
N--anholono\-mic (pseudo) Riemannian space is defined on $C^{\infty }(%
\mathbf{V}^{2n})[[v]]$ by formula
\begin{equation*}
\ ^{1}f\ast \ ^{2}f\doteqdot \sigma (\tau (\ ^{1}f))\circ \sigma (\tau (\
^{2}f)),
\end{equation*}%
where the projection $\sigma :\mathcal{\check{W}}_{^{K}\mathcal{D}%
}\rightarrow C^{\infty }(\mathbf{V}^{2n})[[v]]$ onto the part of $\deg _{s}$%
--degree zero is a bijection and the inverse map $\tau :C^{\infty }(\mathbf{V%
}^{2n})[[v]]\rightarrow \mathcal{\check{W}}_{^{K}\mathcal{D}}$ can be
calculated recursively w.r..t the total degree $Deg,$%
\begin{eqnarray*}
\tau (f)^{(0)} &=&f, \\
\tau (f)^{(k+1)} &=&\ \check{\delta}^{-1}\left( \ ^{K}\mathbf{D}\tau
(f)^{(k)}-\frac{i}{v}\sum\limits_{l=0}^{k}ad_{Wick}(r^{(l+2)})(\tau
(f)^{(k-l)})\right) ,
\end{eqnarray*}%
for $k\geq 0.$
\end{theorem}

Let $\ ^{f}\xi $ be the  Hamiltonian vector field
corresponding to a function $f\in C^{\infty }(\mathbf{V}^{2n})$ on the space $(%
\mathbf{V}^{2n},\check{\theta})$ and consider the antisymmetric
part $$\ ^{-}C(\ ^{1}f,\ ^{2}f)\doteqdot \frac{1}{2}\left( C(\
^{1}f,\ ^{2}f)-C(\ ^{2}f,\ ^{1}f)\right) $$ of a bilinear operator
$C(\ ^{1}f,\ ^{2}f).$ We say that a star--product (\ref{starp}) is
normalized if $\ _{1}C(\ ^{1}f,\ ^{2}f)=\frac{i}{2}\{\ ^{1}f,\
^{2}f\},$ where $\{\cdot ,\cdot \}$ is the Poisson bracket. For  a
normalized $\ast ,$ the bilinear operator $\ _{2}^{-}C$ is a de
Rham--Chevalley 2--cocycle. There is a unique closed 2--form $\
\check{\varkappa}$ such that%
\begin{equation}
\ _{2}C(\ ^{1}f,\ ^{2}f)=\frac{1}{2}\ \check{\varkappa}(\ ^{f_{1}}\xi ,\
^{f_{2}}\xi )  \label{c2}
\end{equation}%
for all $\ ^{1}f,\ ^{2}f\in C^{\infty }(\mathbf{V}^{2n}).$ The class $c_{0}$
of a normalized star--product $\ast $ is defined as the equivalence class $%
c_{0}(\ast )\doteqdot \lbrack \check{\varkappa}].$

A straightforward computation of $\ _{2}C$ from (\ref{c2}) and the
results of Theorem \ref{th2} give the proof of
\begin{lemma}
\label{lem1}The unique 2--form can be computed
\begin{eqnarray*}
\check{\varkappa} &=&-\frac{i}{8}\mathbf{\check{J}}_{\tau }^{\ \alpha
^{\prime }}\ ^{K}\mathbf{R}_{\ \alpha ^{\prime }\alpha \beta }^{\tau }\
\mathbf{\check{e}}^{\alpha }\wedge \ \mathbf{\check{e}}^{\beta }-i\ \check{%
\lambda}, \\
\check{\lambda} &=&d\ \check{\mu},\ \check{\mu}=\frac{1}{6}\mathbf{\check{J}}%
_{\tau }^{\ \alpha ^{\prime }}\ ^{K}\mathbf{T}_{\ \alpha ^{\prime }\beta
}^{\tau }\ \mathbf{\check{e}}^{\beta }.
\end{eqnarray*}
\end{lemma}

Let us define the canonical class $\check{\varepsilon}$ for $\ ^{\check{N}}T%
\mathbf{V}^{2n}=h\mathbf{V}^{2n}\oplus v\mathbf{V}^{2n}$ with the
left label related to a N--connection structure
$\mathbf{\check{N}}.$ The distinguished complexification of such
second order tangent bundles can be performed in
the form $T_{\mathbb{C}}\left( \ ^{\check{N}}T\mathbf{V}^{2n}\right) =T_{%
\mathbb{C}}\left( h\mathbf{V}^{2n}\right) \oplus T_{\mathbb{C}}\left( v%
\mathbf{V}^{2n}\right) .$ In this case, the class $\ \check{\varepsilon}$ is
the first Chern class of the distributions $T_{\mathbb{C}}^{\prime }\left( \
^{N}T\mathbf{V}^{2n}\right) =T_{\mathbb{C}}^{\prime }\left( h\mathbf{V}%
^{2n}\right) \oplus T_{\mathbb{C}}^{\prime }\left( v\mathbf{V}^{2n}\right) $
of couples of vectors of type $(1,0)$ both for the h-- and v--parts.

We calculate the canonical class $\check{\varepsilon}%
,$ using the d--connection $\ ^{K}\mathbf{D}$  considered for
constructing $\ast $ and the h- and v--projections $h\Pi =\frac{1}{2}%
(Id_{h}-iJ_{h})$ and $v\Pi =\frac{1}{2}(Id_{v}-iJ_{v}),$ where $Id_{h}$ and $%
Id_{v}$ are respective identity operators and $J_{h}$ and $J_{v}$ are almost
complex operators, which are projection operators onto corresponding $(1,0)$%
--subspaces. Defining the matrix $\left( h\Pi ,v\Pi \right) \ ^{K}%
\mathbf{R}\left( h\Pi ,v\Pi \right) ^{T},$ where $(...)^{T}$ means
transposition, which is the curvature matrix of the N--adapted restriction
of $\ ^{K}\mathbf{D}$ to $T_{\mathbb{C}}^{\prime }\left( \ ^{N}T\mathbf{V}%
^{2n}\right) ,$ we compute the closed Chern--Weyl form
\begin{eqnarray*}
\check{\gamma} &=& -iTr\left[ \left( h\Pi ,v\Pi \right) \ ^{K}\mathbf{R}%
\left( h\Pi ,v\Pi \right) ^{T}\right] =-iTr\left[ \left( h\Pi ,v\Pi \right)
\ \ ^{K}\mathbf{R}\right] \\
&=& -\frac{1}{4}\mathbf{\check{J}}_{\tau }^{\ \alpha ^{\prime }}\ ^{K}%
\mathbf{R}_{\ \alpha ^{\prime }\alpha \beta }^{\tau }\ \mathbf{\check{e}}%
^{\alpha }\wedge \mathbf{\check{e}}^{\beta }.
\end{eqnarray*}
We get that the canonical class is $\check{\varepsilon}\doteqdot \lbrack
\check{\gamma}].$

\begin{theorem}
The zero--degree cohomology coefficient $c_{0}(\ast )$ for the almost K\"{a}%
hler model of a (pseudo) Riemannian space defined by d--tensor (\ref{hvmetr1}%
) is  given by 
\begin{equation*}
c_{0}(\ast )=-(1/2i)\ \check{\varepsilon}.
\end{equation*}
\end{theorem}

For a particular case when (\ref{hvmetr1}) is a solution of the Einstein
equations, the coefficient $c_{0}(\ast )$ defines certain quantum properties
of the gravitational field. Any metric defining a classical Einstein
manifold can be nonholonomically deformed into the corresponding quantum
configuration.

Finally, in this section we discuss why we have not
considered directly the nonholonomic quantum deformations of
Einstein's equations. In general this is an important subject, 
for further detailed considerations see \cite{vqgr4} 
on explicit constructions with well defined limits, for instance,
in loop gravity and/or any string/ gauge models of gravity.  In
Refs. \cite{vncg1,vncg2},  we re--formulated/ extended the Einstein
equations as Yang--Mills equations for the Cartan connection in
affine / de Sitter frame bundles and considered noncommutative
generalizations of general relativity via Witten--Seiberg maps.
Similar constructions can be elaborated in the nonholonomic
deformation quantization methods of the above-mentioned models. Even 
gauge theories with affine/ Poincare gauge groups are endowed with 
degenerate Killing forms and have an undefined Lagrange structure in the 
total space.  Such theories can be quantized following respective
methods of geometric and/or  BRST quantization
\cite{lyakh1,lyakh2}.

\section{Conclusion}

In this paper  we have provided a natural Fedosov deformation
quantization of (semi) Riemannian spaces equipped with
nonholonomic distributions when the spacetime geometry is
modelled equivalently on almost K\"{a}hler manifolds. This approach
was elaborated following a synthesis of the nonlinear connection
(N--connection) formalism in Lagrange--Finsler geometry and
certain deformation quantization methods.  It also allows to
quantize any solution of Einstein's equations.

In Ref. \cite{vqgr2} we developed a deformation quantization
scheme for general relativity by extending, canonically,  the
geometric structures in extra dimensions or in tangent bundles of 
spacetime manifolds. That approach, in general, 
violates the local Lorentz symmetry which has captured certain
interest in the current  literature.  In this paper the nonholonomic
deformations are considered for the same spacetime manifold which
is being quantized by geometric methods. This allows us to preserve a
formal local Lorentz invariance  even when one works with canonical
distinguished connections which are metric compatible and with
a nonholonomically induced torsion.  All non--quantized and quantized
expressions can be redefined for the Levi Civita connection and this is
possible because all linear connections considered in our approach
are induced uniquely by the metric structure.

It should be emphasized that by prescribing a nonholonomic
distribution,  which defines a nonholonomic frame structure with
an associated nonlinear connection on a (semi) Riemannian manifold,  we
are not violating the general covariance principle. For certain
purposes we only constrain some frame components to be nonintegrable
distributions, but this does not affect the fundamental properties
of physical interactions. In general, all constructions can be
re--defined for arbitrary frame and coordinate systems.

Any $2+2$ splitting of an Einstein manifold can be performed in a
natural nonholonomic fashion in order to generate  the required almost symplectic
structures and apply, straightforwardly,  the deformation
quantization methods.  In this way we deform both the frame and
linear connections structures in a canonical fashion
when the constructions are defined in terms of the metric structure (in
particular, by a solution of the Einstein equations).

The study of nonholonomic geometry and deformation quantization of
gravity in connection to Loop Quantum Gravity,  canonical and perturbative
approaches, noncommutative generalizations and applications to
modern cosmology and gravitational physics will be the subject of 
future investigation.

\vskip3pt

\textbf{Acknowledgement: } The work is performed during a visit at
the Fields Institute. The author is grateful to C. Castro Perelman
for discussions.

\end{document}